\documentclass[iop]{emulateapj}
\usepackage{hyperref}

\begin{document}

%% one line. However, you may use \\ to force a line break if you
%% desire.

\title{Tails from the Orphanage}

\author{Carl J. Grillmair}
\affil{IPAC, Mail Code 314-6, Caltech, 1200 E. California Blvd., Pasadena, CA 91125}
\email{carl@ipac.caltech.edu}

\begin{abstract}

Examining a portion of the northern Sloan Digital Sky Survey (SDSS) 
footprint, we detect at least three and possibly seven halo debris
streams. One of these (PS1-D) was recently detected in the Pan-STARRS1
$3\pi$ survey, and the remaining two are also evident as extensions of
the SDSS detections. All of these streams are metal poor and are
found at a distance of around $21 \pm 5$ kpc. The streams are between
65\arcdeg~ and 70\arcdeg~ in length, oriented almost north-south, and
are nearly parallel and somewhat convergent with the neighboring
Orphan stream. Surface densities ranging from 1.5 to 0.5 stars per
square degree down to $g = 21.7$ correspond to surface brightnesses
between 35 and 37 mag per square arcsecond. The streams each
appear to be more than 300 pc across, suggesting either 
dwarf/ultrafaint galaxy progenitors or long-term heating of very
ancient globular cluster streams. The orbits of all but one of these
streams appear to be nearly radial, and the orbit normals suggest that
all of the streams are part of the Vast Polar Structure, a relatively
narrow plane that contains most of the known satellite galaxies,
globular clusters, and stellar streams.

\end{abstract}



\section{Introduction}

Dozens of distinct, highly collimated stellar debris streams are now
known to orbit in the halo of our Galaxy (see \citet{grillmair2016}
and \citet{smith2016} for reviews). Detecting and tracing such streams
around the Galaxy will become increasingly important as we refine our
techniques for using them as probes of the Galactic potential
\citep{kupper2015, bovy2016} and of the distribution of dark matter subhalos
\citep{carlberg2009, yoon2011}.  Knowing in advance the locations and
trajectories of these streams may also help us interpret the vast
amount of data we expect to harvest from upcoming Gaia data
releases.

While searching for and examining other streams in the Sloan Digital
Sky Survey (SDSS), we have long noted a somewhat fibrous texture in
the area of sky surrounding the Orphan stream (the
``orphanage''). However, detailed examinations at high spatial
resolution (0.1\arcdeg) have not revealed any obvious streams at a
signal-to-noise ratio sufficient for publication. In this Letter we
coarsen our spatial sampling considerably, enabling the detection of at
least three nearly parallel, low-metallicity stellar debris
streams. We describe our detection method in Section
\ref{analysis}. We make an initial attempt to constrain the orbits in
Section \ref{orbits}, and discuss additional low-significance
structures in Section \ref{otherstreams}. Concluding remarks are given
in Section \ref{conclusions}.

\section{Analysis} \label{analysis}

We make use of the photometric catalog from data release 10 of the
SDSS. Experience has shown that we probe most deeply and are most
sensitive to differences in stellar populations using just the $g, r,$
and $i$ measurements. We use all objects classed as stars and with $g
< 21.7$. Photometry is dereddened using the DIRBE/IRAS dust maps of
\citet{schleg98}, corrected using the prescription of
\citet{schlafly2011}. 

Figure 1 shows the result of applying a matched filter in the
color-magnitude domain to the western half of the northern footprint
of the SDSS. The filter is based on the color-magnitude distribution
of stars in the old, metal-poor globular cluster NGC 5053 with [Fe/H]
$\approx -2.29$ \citep{harris1996}. Three streams become apparent,
running roughly north-south and somewhat convergent with the Orphan
stream, similar in appearance to the claw marks that bears commonly
leave on trees.

Figure 2 shows the distribution of $E(B-V)$ over the same region of
sky as in Figure 1 \citep{schleg98} . While there are evidently some
random clumps of dust emission between the streams, there are no
linear features that resemble the streams in Figure 1. We conclude
that the features in Figure 1 are not an artifact of reddening-induced
incompleteness.

Figure 3 shows the same part of the sky in a matched-filtered map of
the Pan-STARRS $3\pi$ survey \citep{bernard2016} for a distance of 25
kpc. While there are evidently issues with spatially variable
completeness at the limit of the single-epoch survey, each of the
streams in Figure 1 has a counterpart in Figure 3 extending to the
southern limit of the Pan-STARRS survey. That the streams are not as obvious as
they are in Figure 1 is presumably at least partly due to the more
metal-rich ([Fe/H] = -1.5) isochrone used by
\citet{bernard2016}. For example, the Orphan stream
appears considerably weaker in Figure 3 than it does in
\citet{grill2006a}, \citet{belokurov2006}, or \citet{newberg2010}.

The features in Figures 1 and 3 extend through the 
constellations Leo, Leo Minor, Hydra, Sextans, Antlia, and Pyxis.
Since there are also multiple streams in the same constellations, we follow
\citet{grillmair2014} and name the streams after rivers cited in 
the Illiad. Henceforth we refer to the stream next to PS1-D as 
Sangarius, and the stream next to Orphan as Scamander.

The trajectories of the streams north of the bright, southern
Sagittarius arm in panel d of Figure 1 are obviously somewhat conjectural. While
they appear plausible, the discontinuity created by the Sagittarius
stream creates some ambiguity, and velocity information will be
required before we can definitively conclude that the northern
extensions are not unrelated structures. Including both northern
(Figure 1) and southern portions (Figure 3) of the streams, the arc
lengths are 68\arcdeg, 59\arcdeg, and 66\arcdeg~ for PS1-D, Sangarius,
and Scamander, respectively. At a distance of 21 kpc, this translates
to physical lengths of 25, 22, and 24 kpc, respectively.

In equatorial coordinates, the trajectory of PS1-D can be modeled
to $\sigma = 0.19\arcdeg$ using:

\begin{eqnarray}
\alpha = 141.017 + 0.208 \delta - 0.02491 \delta^2 + 0.000609 \delta ^3
\nonumber \\
- 1.20989 \times 10^{-6} \delta^4
\end{eqnarray}

\noindent Sangarius is well modeled ($\sigma = 0.22\arcdeg$) with:

\begin{equation}
\label{sangarius}
\alpha = 148.9492 - 0.03811 \delta + 0.001505 \delta^2 
\end{equation}

\noindent while Scamander again requires a higher-order fit ($\sigma = 0.11\arcdeg$):

\begin{eqnarray}
\label{sangarius}
\alpha = 155.642 - 0.1000 \delta - 0.00191 \delta^2 - 0.0003346 \delta
^3 \nonumber \\
+ 1.47775 \times 10^{-5} \delta^4
\end{eqnarray}

\citet{grill2009} quantified the significance of streams above the
background using the ``T'' statistic, measuring the median filtered
surface density in multiple segments as one sweeps across the sky
perpendicular to the orientation of the stream. The region of sky in
the vicinity of the orphanage is quite complex, with the Sagittarius,
Orphan, EBS, and AntiCenter streams, along with the orphanage streams
themselves, making it difficult to measure the noise floor in a
stream-free region. In Figure 4 we show the run of {\it T} across an
unsmoothed version of Figure 1, normalizing to the ``field'' RMS
measured in an identical manner in an apparently blank region of sky
to the north and east of the Sagittarius stream. We use equations 1,
2, and 3 to define the stream segments we pass over each respective
stream. We detect PS1-D and Scamander at roughly the
$5\sigma$ level, while we detect Sangarius at somewhere between 2
and $4\sigma$. These levels roughly accord with the visual impression
in Figure 1.

Sampling a version of Figure 1 using $0.1\arcdeg$ binning, we find
that the FWHMs of the streams range from
0.9\arcdeg~ to 1.1\arcdeg. At a distance of 21 kpc, this corresponds
to physical widths ranging from 330 to 400 pc, broader than most of
the known globular cluster streams but narrower than the Orphan
stream. This suggests that the streams could be either the remnants of
diminutive dwarf or ultrafaint galaxies, or (more likely given our
population estimates below) globular cluster streams that have been
heated for 10 Gyr or more by dark-matter subhalos
\citep{carlberg2009}.

Color-magnitude diagrams for the streams are shown in Figure 5. All of
the streams appear to be quite metal poor ([Fe/H] $\le -1.4$), which is
expected since our filter was designed to capture metal-poor
substructures. To estimate the number of stars in each stream, we
count only stars that lie within $2\sigma$ of the NGC 5053
color-magnitude locus and on the subgiant branch or below. Once again,
given the complexity of the field, it is difficult to measure the
surface density of field stars. Our best estimate yields mean stream
surface densities of 0.5 stars per square degree for the southernmost
$15\arcdeg$ ($-35\arcdeg < \eta < -20\arcdeg $) of PS1-D and Sangarius
in Figure 1 and 1.5 stars per square degree for the same portion 
of Scamander. This is superposed on a field star surface density,
selected using the same photometric constraints, of about 10 stars per
square degree. We are therefore sampling the streams with a total of
between 30 and 110 stars down to $g = 21.7$. If we adopt a globular
cluster-like luminosity function and integrate down the main sequence,
then for a distance of 21 kpc, we find that there should be between
440 (Sangarius) and 1600 (Scamander) stars in total within these
portions of the streams. A similar integration yields equivalent
surface brightnesses of between 35 and 37 magnitudes per square
arcsecond. If the stream segments we see are portions of streams that
encircle the Galaxy, then we arrive at total populations of the
progenitors of between $10^4$ and $4 \times 10^4$ stars,
e.g. comparable to the populations of modern-day globular clusters.

There are no known globular clusters or dwarf galaxies falling near
great circles defined by each stream. though this is not conclusive as the
uncertainties are high. On the other hand, the widths of the streams,
combined with their tenuousness, may be an indication of great age
and of the possibility that the progenitors dissolved long ago.

\section{Orbits} \label{orbits}

Though we have as yet no velocity information for these streams, and
while we are mindful of the fact that streams do not precisely follow
single orbits \citep{eyre2011}, the distances and trajectories of the
streams can give us some idea of the orbit parameters. We use a model
for the Galactic potential by \citet{allen1991} and fit in a
least-squares sense to five or six, roughly equidistant normal points
along each stream in the region $-35\arcdeg < \eta < 0\arcdeg$. We
assign a distance of 21 kpc to the middle and the ends of each
stream. While our distance estimates are far less certain than the
stream trajectories on the sky, we find that the ends of the streams
disappear if we shift the selection filter by more than $\approx 0.5$
mag faintward or brightward; the streams are evidently nearly
perpendicular to our line of sight. We assign $0.3\arcdeg$
uncertainties to the positions of the individual normal points and 5
kpc uncertainties to our distance estimates.

Table 1 lists the orbit parameters corresponding to the best fits. The
uncertainties are estimated by setting each of the three free
parameters (the radial velocity and the two components of proper
motion) to their 90\% limits and refitting the stream using the
remaining two parameters. While the orbital planes are quite well
constrained by the observed trajectories of the streams, the radial
velocities and hence eccentricities are evidently nearly
unconstrained.

The convex-eastwards curvature found by \citet{bernard2016} for PS1-D
is at odds with any plausible orbit around the Galaxy. This
westward-bending feature at the north end of the stream is visible in
panels (b) and (c) of Figure 1, though in the SDSS data it is neither
an obvious continuation of the stream nor clearly preferred over the
eastward-bending feature we trace in panel (d). The stream trace in
panel (d) alleviates the nonphysical aspect of the orbit, though only
just.  This suggests that the northern part of the Bernard et al. (2016)
stream may actually be some unrelated structure. Deeper photometric
data and/or Gaia proper motions will be required to determine the true
path of PS1-D north of the Sagittarius stream.

Both our eastward-bending PS1-D and Sangarius appear to be on nearly
radial orbits, while Scamander is on a less extreme, somewhat more
tightly bound orbit. We note that the Orphan stream is also on a
far-flung orbit, with $R_{apo} \simeq 90$ kpc \citep{newberg2010}. 
This suggests the possibility that PS1-D and Sangarius may be dynamically
related to Orphan in some way, perhaps through a much larger structure
that gave rise to Orphan. Table 1 shows that the orbital poles for
PS1-D, Sangarius, and Scamander all lie between those of the Orphan
and Anticenter streams. This consequently puts them within the region
defining the Vast Polar Structure (VPOS) \citep{pawlowski2012} that
apparently contains the orbits of most of the known dwarf galaxies,
globular clusters, and stellar debris streams orbiting our Galaxy.

\section{Still Other Streams?} \label{otherstreams}

Careful examination of Figure 1 shows that there may be at least four
more streams, at S/Ns considerably lower than those
of the streams analyzed above, at a distance of $\approx 21$ kpc in
the region $0\arcdeg > \lambda > -25\arcdeg, -5\arcdeg < \eta <
25\arcdeg$. These ladder-like structures are almost certainly affected
by discontinuities in the SDSS scan direction, but the enhancements in
the north-south direction are less easy to dismiss. 

If indeed these are additional streams, then their similar
orientations suggest that either the data or our filtering and
analysis are somehow particularly sensitive to streams oriented in a
nearly north-south direction, or that the VPOS once contained many
more objects (clusters or dwarf galaxies) than we see today.
Concerning the former, we are certainly biased {\it against} selecting
structures oriented along the SDSS east-west scan direction, but we
know of no process or peculiarity that could enhance apparently linear
structures in the cross-scan direction. These potential streams will
need to be verified once we have access to other surveys
(e.g. Pan-STARRS) and other stream detection methods (e.g. Gaia). We
mention them here primarily to aid in the identification or
verification of structures that may be discovered in upcoming Gaia
releases.

These features appear to have trajectories qualitatively similar to those
of the three streams examined above. Orbit fits yield nearly radial
orbits, with orbit normals ($l, b = 199\arcdeg \pm 1\arcdeg,
26\arcdeg \pm 4\arcdeg$) that once again lie within the patch of sky
associated with the VPOS.

\section{Conclusions} \label{conclusions}

Examining a region of the northern footprint of the Sloan Digital Sky
Survey, we find evidence for at least three, and possibly seven
stellar debris streams. With 30 to 110 stars per stream and equivalent
surface brightnesses $> 35$ mag arcsec$^{-2}$, these streams are likely
at the very limit of what can be detected in the SDSS. Yet
they are enticing as perhaps the tip a substantial iceberg, as well as an
indicator of what we may discover when Gaia proper motions become
available.

All of these streams appear to orbit within the VPOS,
believed to contain the majority of surviving satellite galaxies. Once 
this has beenverified with velocity information, these streams will clearly add to the
significance of this structure and to the consequences it may have for
our understanding of galaxy formation in $\Lambda$CDM.

\begin{acknowledgements}

We are grateful to Edouard Bernard for making available
matched-filtered maps of the Pan-STARRS $3\pi$ survey. We are also
grateful to an anonymous referee for several thoughtful suggestions
that improved both the content and readability of the paper.

\end{acknowledgements}

{\it Facilities:} \facility{Sloan, PS1}

\clearpage

\begin{figure}
\epsscale{1.0}
\plotone{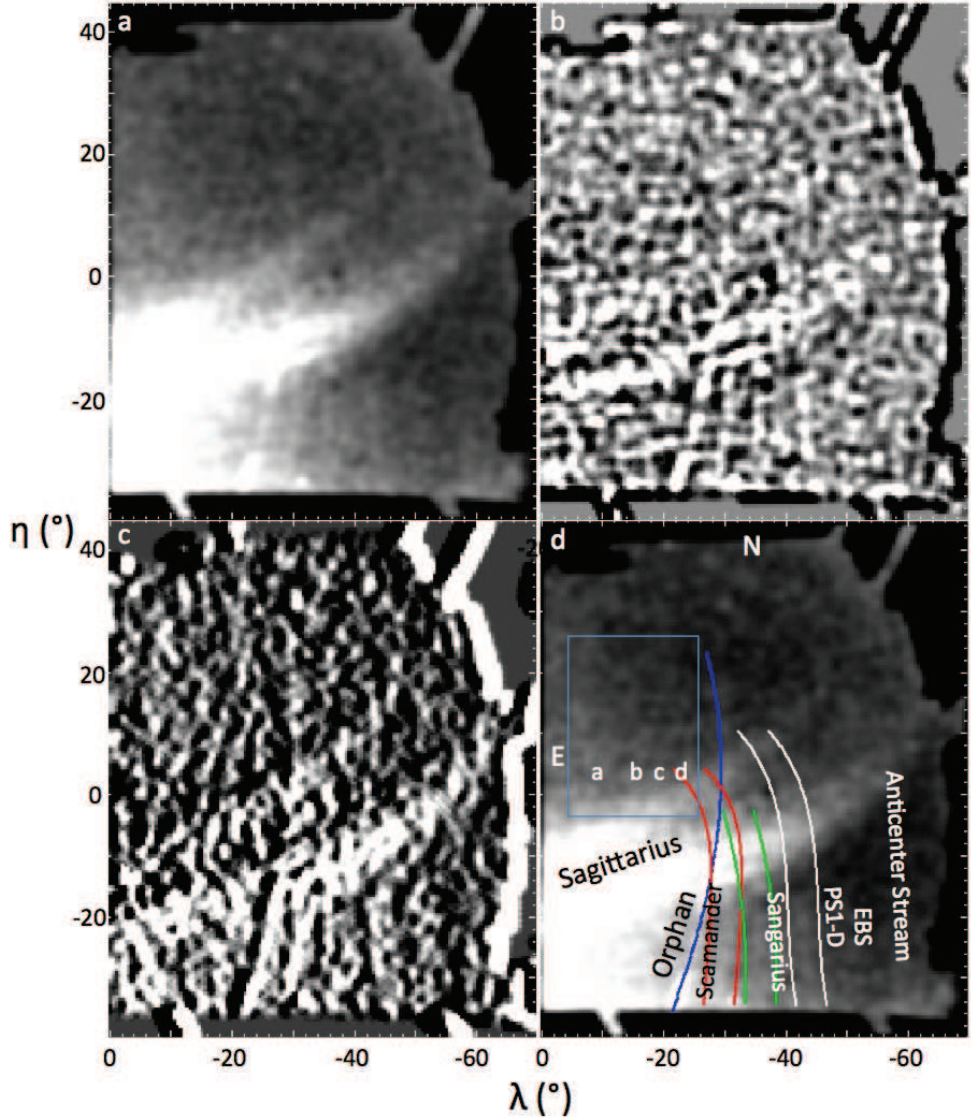}
\caption{Panel (a): filtered surface density map of the $70\arcdeg
  \times 85\arcdeg$ western portion of the northern SDSS footprint, in
  SDSS coordinates. The SDSS coordinate system has the advantage that
  small differences in calibration and completeness from scan to scan
  are aligned east-west, or very nearly along horizontal lines in the Figure.
  The matched filter is based on the color-magnitude distribution of
  stars in NGC 5053, shifted to a distance of 21 kpc. The map has been
  binned to $0.5\arcdeg\times0.5\arcdeg$ pixels, and smoothed with a
  Gaussian kernel of 1.5\arcdeg. The stretch is linear, with lighter
  areas indicating higher surface densities. Panel (b): the same
  filtered image as in panel (a) after subtraction of a smoothed
  background image constructed using an annular median window filter
  of radius 4.5\arcdeg. Panel (c): The map in panel (a) after
  subtracting a version of itself that is shifted 1\arcdeg~ to the
  east. Panel (d): same as panel (a), but with the streams indicated. The
  curves correspond to Equations 1, 2, and 3, shifted east and west by
  2.5\arcdeg. The blue curve shows the Orphan stream trajectory of
  \citet{newberg2010}. The light blue box indicates the location of
  four possible streams (labeled a, b, c, and d) discussed in Section
  \ref{otherstreams}.}
\end{figure}

\begin{figure}
\epsscale{1.0}
\plotone{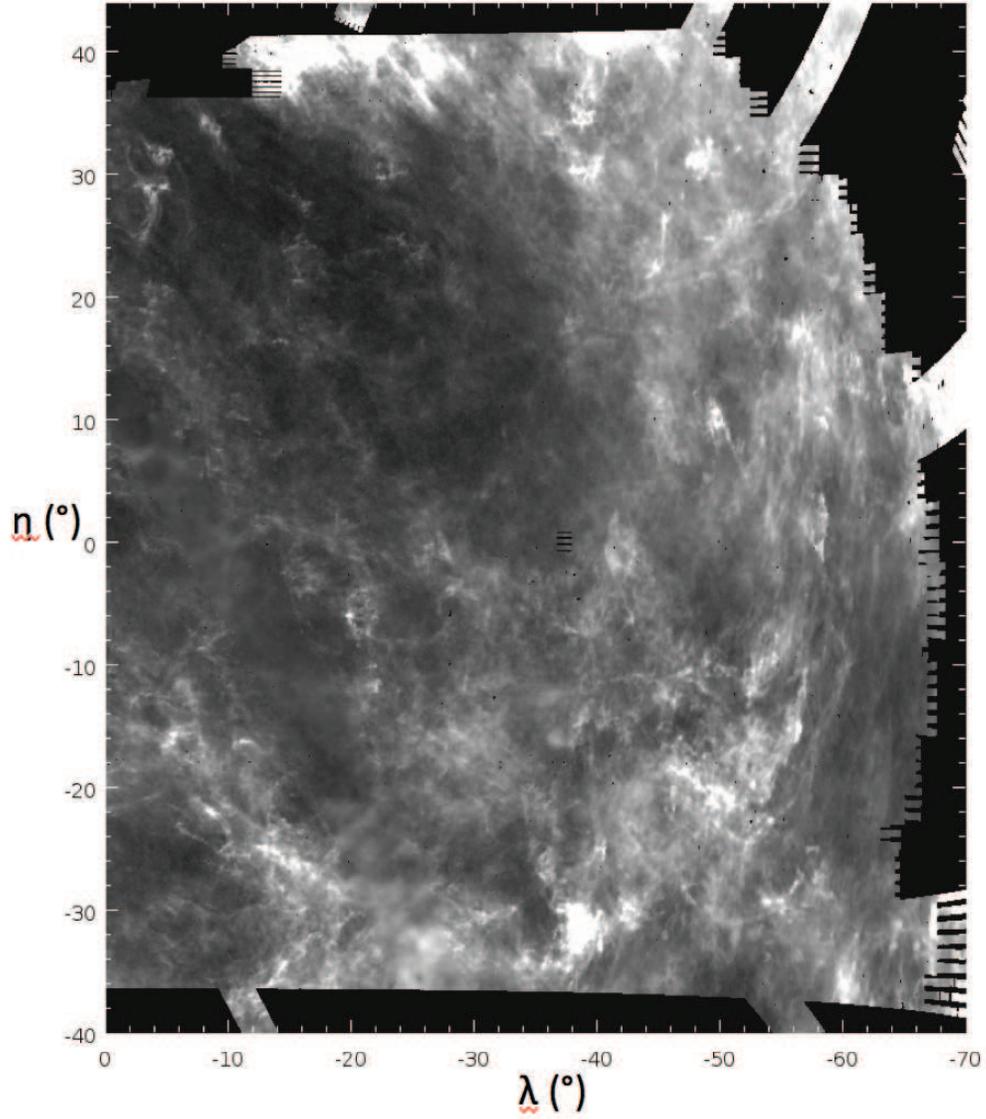}
\caption{The distribution of $E(B-V)$, taken from \citet{schleg98}, in
  the same coordinate system as Figure 1. The stretch is linear, with
  white indicating $E(B-V) > 0.09$ and black corresponding to $E(B-V)
  < 0.01$}
\end{figure}

\begin{figure}
\epsscale{1.0}
\plotone{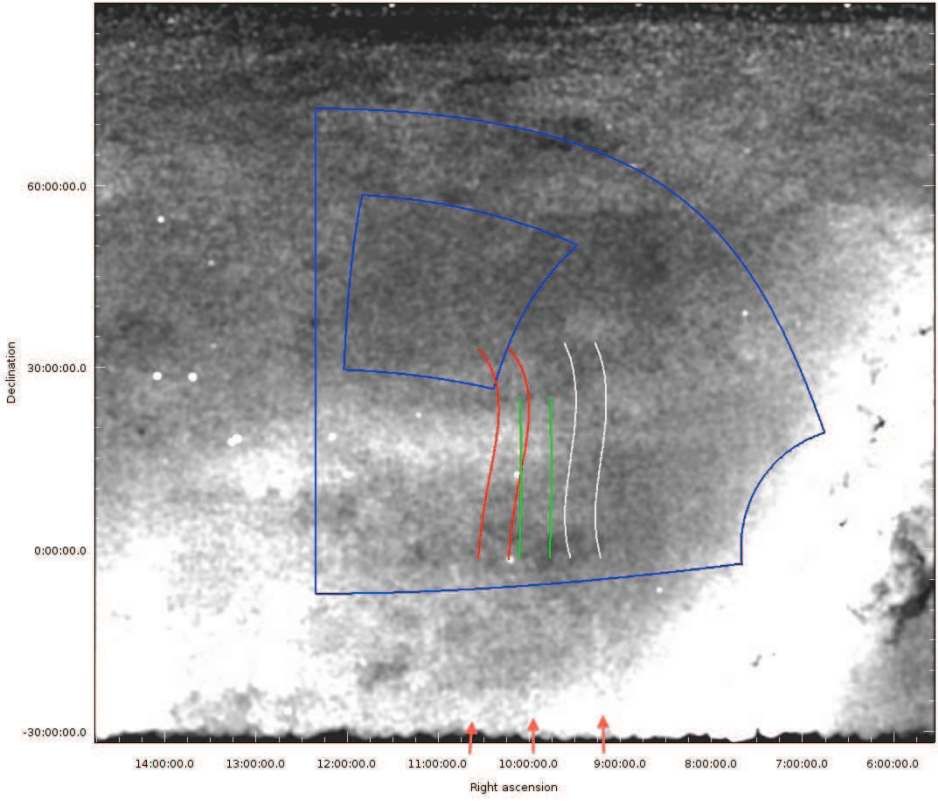}
\caption{A portion of Bernard et al.'s (2016) matched-filtered density
  map of stars in a single epoch of the Pan-STARRS $3\pi$ survey. In
  this case, the stars have been filtered using an isochrone with an
  age of 12 Gyr, a metallicity of [Fe/H] = -1.5, and a distance of 25
  kpc. The blue boxes indicate the region shown in Figure 1, and the
  red, green, and white curves are the same as those in panel d of
  Figure 1.  The stretch is linear, with lighter areas indicating
  higher surface densities. Despite the significant pattern noise due
  to variations in depth and completeness, Sangarius, Scamander, and
  PS1-D all show indications of continuing southward, as indicated by the
  arrows.}

\end{figure}

\begin{figure}
\epsscale{1.0}
\plotone{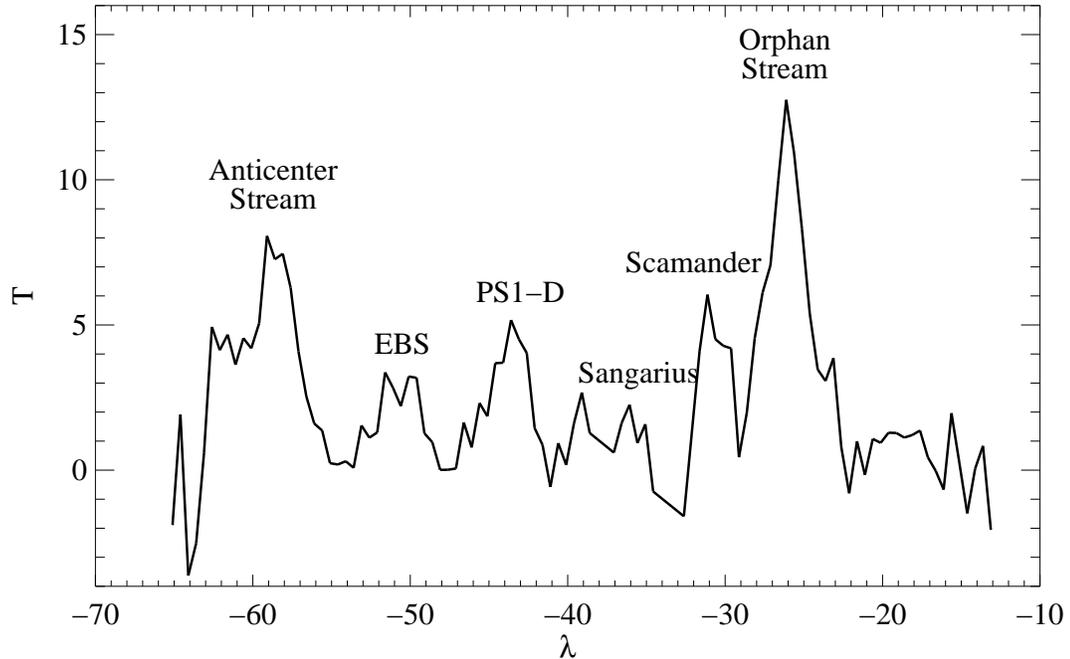}
\caption{The ``T'' statistic of \citet{grill2009} for the southern
  15\arcdeg~ of the streams in Figure 1. {\it T} is the median value,
  measured from the image filtered for a distance of 21 kpc, for five,
  3\arcdeg long segments in each stream. We have normalized the run of
  {\it T} by dividing by the RMS measured the same way in an apparently
  blank region of sky north of the Sagittarius streams. The plotted
  values thus correspond roughly to the S/N at each
  point. The peaks corresponding to the Anticenter Stream
  \citep{grill2006b, li2012}, EBS \citep{grill2011, hargis2016}, and
  Orphan streams \citep{grill2006a, belokurov2007, newberg2010,
    sesar2013, grillmair2015} show up only incidentally and are
  significantly lower than the S/Ns we would measure
  using more appropriate filters, correct distances, and suitably
  oriented stream segments.}
\end{figure}

\begin{figure}
\epsscale{1.0}
\plotone{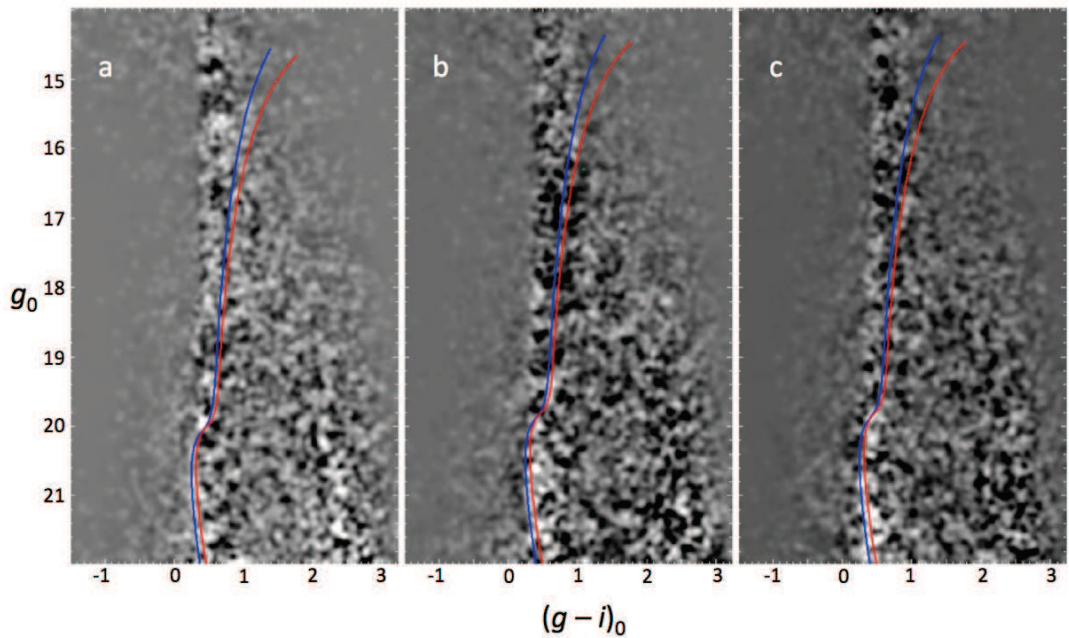}
\caption{Dereddened $(g -i)_0$ Hess diagrams for (a) PS1-D, (b) Sangarius, and (c) Scamander. The curves show the theoretical loci for Z = 0.0001 ([Fe/H] = -2.2) in blue and Z = 0.0007 ([Fe/H] = -1.44) in red.}
\end{figure}

%\floattable
\begin{deluxetable}{clcccc}
%\rotate
\tabletypesize{\small}
\tablecaption{Predicted Motions and Orbit Parameters}
\tablecolumns{5}
%\tablewidth{0pt}

\tablehead{& & \colhead{PS1-D} & \colhead{Sangarius} & \colhead{Scamander}} \\
\startdata
Fiducial Point & R.A. ($^\circ$, J2000) & 141.457 & 149.013 & 153.520 \\
\\
& dec ($^\circ$, J2000) & +11.301 & +7.847 & +12.460 \\
\hline \\
& $v_{hel}$ (km s$^{-1}$) & $63 \pm 109 $ & $156^{+314}_{-421}$ & $151^{+38}_{-46}$ \\
Prograde Orbit & $\mu_\alpha$ cos($\delta$) (mas yr$^{-1}$) & $-0.31 \pm 0.06$ & $-0.78 \pm 0.19$ & $-0.47 \pm 0.02$ \\
& $\mu_\delta$ (mas yr$^{-1}$) & $+3.16 \pm 0.9$ &
$+6.21^{+2.0}_{-1.0}$ & $-1.01 \pm 0.07$\\
\\
\hline \\
& $v_{hel}$ (km s$^{-1}$) &  $159 \pm 109$ & $84^{+421}_{-314}$ & $ 53^{+46}_{-38}$ \\
Retrograde Orbit & $\mu_\alpha$ cos($\delta$) (mas yr$^{-1}$) & $+0.14 \pm 0.06$ & $+0.21 \pm 0.19$ & $-0.25 \pm 0.02$\\
& $\mu_\delta$ (mas yr$^{-1}$) & $-7.0 \pm 0.9 $ &
$-9.89^{+2.0}_{-1.0} $ & $-2.55 \pm 0.07 $ \\
\\
\hline \\
$R_{apo}$ (kpc) && $> 100$ & $> 100 $ & $29 \pm 2 $ \\
$R_{peri}$ (kpc) && $27 \pm 0.3$ & $25 \pm 0.2$ & $6 \pm 1$  \\
$i$ ($^\circ$) && $64 \pm 3$ & $79 \pm 5$ & $34 \pm 1$ \\
$\epsilon$ && $ >0.6 \pm 0.3$ & $>0.6 \pm 0.3$ & $ 0.66 \pm 0.03$ \\
Orbit Pole & $l$ ($^\circ$) & $183 \pm 2$ & $178 \pm 5 $& $ 196 \pm 2$ \\
&            $b$ ($^\circ$) & $-57 \pm 1$ & $-52 \pm 1 $& $-41 \pm 2 $ \\
\enddata
\end{deluxetable}

\end{document}